\documentclass[twocolumn,aps,prl,showpacs]{revtex4}
\usepackage[T1]{fontenc}
\usepackage[latin9]{inputenc}
\usepackage{amstext}
\usepackage{amssymb}
\usepackage{graphicx}
\usepackage{esint}

\makeatletter
\@ifundefined{textcolor}{}
{%
 \definecolor{BLACK}{gray}{0}
 \definecolor{WHITE}{gray}{1}
 \definecolor{RED}{rgb}{1,0,0}
 \definecolor{GREEN}{rgb}{0,1,0}
 \definecolor{BLUE}{rgb}{0,0,1}
 \definecolor{CYAN}{cmyk}{1,0,0,0}
 \definecolor{MAGENTA}{cmyk}{0,1,0,0}
 \definecolor{YELLOW}{cmyk}{0,0,1,0}
}

\makeatother

\begin{document}

\title{Universal impurity-induced bound state in topological superfluids}

\author{Hui Hu$^{1}$, Lei Jiang$^{2}$, Han Pu$^{3}$, Yan Chen$^{4}$,
and Xia-Ji Liu$^{1}$}

\email{xiajiliu@swin.edu.au}

\affiliation{$^{1}$Centre for Atom Optics and Ultrafast Spectroscopy, Swinburne
University of Technology, Melbourne 3122, Australia \\
 $^{2}$Joint Quantum Institute, University of Maryland and National
Institute of Standards and Technology, Gaithersburg, Maryland 20899,
USA \\
 $^{3}$Department of Physics and Astronomy, and Rice Quantum Institute,
Rice University, Houston, TX 77251, USA \\
 $^{4}$Department of Physics, State Key Laboratory of Surface Physics
and Laboratory of Advanced Materials, Fudan University, Shanghai,
200433, China }
\begin{abstract}
We predict a universal mid-gap bound state in topological superfluids,
induced by either non-magnetic or magnetic impurities in the strong
scattering limit. This universal state is similar to the lowest-energy
Caroli-de Gennes-Martricon bound state in a vortex core, but is bound
to localized impurities. We argue that the observation of such a universal
bound state can be a clear signature for identifying topological superfluids.
We theoretically examine our argument for a spin-orbit coupled ultracold
atomic Fermi gas trapped in a two-dimensional harmonic potential,
by performing extensive self-consistent calculations within the mean-field
Bogoliubov-de Gennes theory. A realistic scenario for observing universal
bound state in ultracold $^{40}$K atoms is proposed. 
\end{abstract}


\maketitle
Topological superfluids are of great interest \cite{Qi2011}. They
are promising candidates that host Majorana fermions \cite{Majorana1937},
which lie at the heart of topological quantum information and computation,
due to their exotic non-Abelian exchange statistics \cite{Kitaev2006,Nayak2008,Wilczek2009}.
To date, there are a number of proposals for practical realizations
of topological superfluids, including $p+ip$ superconductors \cite{Read2000,Mizushima2008},
surfaces of three-dimensional (3D) topological insulators \cite{Fu2008,Sau2010,Alicea2010}
or one-dimensional (1D) spin-orbit coupled nanowires \cite{Lutchyn2010,Oreg2010}
in proximity to an $s$-wave superconductor, and two-dimensional (2D)
\cite{Zhang2008,Sato2009,Zhu2011,Liu2012a} or 1D \cite{Jiang2011,Liu2012b,Wei2012}
spin-orbit coupled atomic Fermi gases near Feshbach resonances. All
these proposals are appealing and are to be examined experimentally.
In fact, recent experimental results on tunneling spectroscopy of
semiconductor InSb nanowires in a magnetic field placed in contact
with a superconducting electrode \cite{Mourik2012} may already suggest
the existence of topological superfluids and Majorana fermions. However,
{\em unambiguous} characterizations of topological properties of
the nanowires are still missing.

In this Letter, we propose that a {\em universal} mid-gap bound
state, induced by strong non-magnetic or magnetic impurity scattering,
could provide a clear signature for the existence of topological superfluids.
In solid state, impurities are widely known to serve as an important
local probe that characterizes the quantum state of hosting systems
\cite{Balatsky2006}. Individual impurities have used to determine
the superconducting pairing symmetry of unconventional non-\textit{s}-wave
superconductors \cite{Mackenzie1998} and to demonstrate Friedel oscillations
on Be(0001) surface \cite{Sprunger1997}. In strongly-correlated many-body
systems, they may be employed to pin one of the competing orders \cite{Millis2003}.
Here, unique to topological superfluids, we predict that a single
impurity with sufficiently strong scattering strength can create a
universal mid-gap state bound to the impurity. It resembles the lowest-energy
Caroli-de Gennes-Martricon (CdGM) bound state inside a vortex core
\cite{CdGM}. For small order parameters, where the bound state energy
$E$ is nearly zero, the wave-function of the universal bound state
is found to closely follow the symmetry of that of Majorana fermions
\cite{Liu2012a}.

In our work, the emergence of universal impurity-induced bound state
is examined theoretically in an interacting spin-orbit coupled ultracold
atomic Fermi gas in 2D harmonic traps \cite{Liu2012a}. We perform
numerically extensive self-consistent calculations by using fully
microscopic Bogoliubov-de Gennes (BdG) theory, to explore the details
of the universal bound state. This specific choice of topological
superfluids is motivated by the recent realization of spin-orbit coupling
in atomic Fermi gases of $^{40}$K \cite{exptShanXi} and $^{6}$Li
atoms \cite{exptMIT}. Benefited from the high controllability in
interaction, geometry and purity in cold-atom experiments, 2D spin-orbit
coupled atomic Fermi gases are arguably the best candidate for observing
the predicted universal bound state. Our results, however, should
be applicable as well to various topological superfluids that are
believed to exist in solid state. We propose a realistic scenario
of creating universal bound state in $^{40}$K atoms and discuss briefly
the relevance of our results to other solid state systems.

\textit{Mean-field BdG equation. }--- To start, we consider a trapped
2D atomic Fermi gas with a Rashba-type spin-orbit coupling and a Zeeman
field $h$, which is believed to be a topological superfluid when
the Zeeman field exceeds a threshold $h_{c}$ \cite{Liu2012a}. The
model Hamiltonian of the system is given by, ${\cal H}=\int d{\bf r}[{\cal H}_{0}({\bf r})+{\cal H}_{I}({\bf r})+{\cal H}_{{\rm imp}}({\bf r})]$,
where 
\begin{equation}
{\cal H}_{0}({\bf r})=\sum_{\sigma=\uparrow,\downarrow}\psi_{\sigma}^{\dagger}{\cal H}_{\sigma}^{S}({\bf r})\psi_{\sigma}+\left[\psi_{\uparrow}^{\dagger}V_{SO}({\bf r})\psi_{\downarrow}+\text{H.c.}\right]
\end{equation}
is the single-particle Hamiltonian density in the presence of Rashba
spin-orbit coupling $V_{SO}({\bf r})=-i\lambda(\partial_{y}+i\partial_{x})$,
${\cal H}_{I}({\bf r})=U_{0}\psi_{\uparrow}^{\dagger}({\bf r})\psi_{\downarrow}^{\dagger}({\bf r})\psi_{\downarrow}({\bf r})\psi_{\uparrow}({\bf r})$
represents the interaction, and ${\cal H}_{{\rm imp}}({\bf r})=\sum_{\sigma=\uparrow,\downarrow}\psi_{\sigma}^{\dagger}V_{{\rm imp}}^{\sigma}({\bf r})\psi_{\sigma}$
describes the potential scattering due to the impurity. Here, $\psi_{\uparrow,\downarrow}^{\dagger}$
are respectively the creation field operators for the spin-up and
spin-down atoms and, ${\cal H}_{\sigma}^{S}({\bf r})\equiv-\hbar^{2}\nabla^{2}/(2M)+M\omega^{2}r^{2}/2-\mu-h\sigma_{z}$
is the single-particle Hamiltonian in a 2D harmonic trapping potential
$M\omega^{2}r^{2}/2$, in reference to the chemical potential $\mu$.
We have used the standard $s$-wave contact interaction between atoms
with opposite spins, whose strength $U_{0}$ is to be regularized
by the binding energy of the two-body bound state $E_{a}$ \cite{Liu2012a,Randeria1989}.
For computational simplicity, we place an impurity at origin and consider
either a delta-like scattering potential, $V_{{\rm imp}}^{\sigma}({\bf r})=V_{{\rm imp}}^{\sigma}\delta(r)$,
or a gaussian-shape potential with width $d$, $V_{{\rm imp}}^{\sigma}({\bf r})=[V_{{\rm imp}}^{\sigma}/(\pi d^{2})]\exp[-r^{2}/d^{2}]$.
In the case of magnetic impurity, we take the potential strength $V_{{\rm imp}}^{\uparrow}=-V_{{\rm imp}}^{\downarrow}=-V_{{\rm imp}}$;
while for non-magnetic impurity, $V_{{\rm imp}}^{\uparrow}=$ $V_{{\rm imp}}^{\downarrow}=-V_{{\rm imp}}$.
We have checked both positive and negative values of $V_{{\rm imp}}$
and have observed very similar results at large $\left|V_{{\rm imp}}\right|$.
Hereafter, we focus on the case with $V_{{\rm imp}}>0$.

We solve the low-energy fermionic quasiparticles of the model Hamiltonian
by using the standard mean-field BdG approach, ${\cal H}_{{\rm BdG}}\Psi_{\eta}\left({\bf r}\right)=E_{\eta}\Psi_{\eta}\left({\bf r}\right)$,
where \begin{widetext} 
\begin{equation}
{\cal H}_{{\rm BdG}}=\left[\begin{array}{cccc}
{\cal H}_{\uparrow}^{S}({\bf r})+V_{{\rm imp}}^{\uparrow}({\bf r}) & V_{SO}({\bf r}) & 0 & -\Delta({\bf r})\\
V_{SO}^{\dagger}({\bf r}) & {\cal H}_{\downarrow}^{S}({\bf r})+V_{{\rm imp}}^{\downarrow}({\bf r}) & \Delta({\bf r}) & 0\\
0 & \Delta^{*}({\bf r}) & -{\cal H}_{\uparrow}^{S}({\bf r})-V_{{\rm imp}}^{\uparrow}({\bf r}) & V_{SO}^{\dagger}({\bf r})\\
-\Delta^{*}({\bf r}) & 0 & V_{SO}({\bf r}) & -{\cal H}_{\downarrow}^{S}({\bf r})-V_{{\rm imp}}^{\downarrow}({\bf r})
\end{array}\right]
\end{equation}
\end{widetext}is the BdG Hamiltonian, $\Psi_{\eta}\left({\bf r}\right)=[u_{\uparrow\eta},u_{\downarrow\eta},v_{\uparrow\eta},v_{\downarrow\eta}]^{T}$
and $E_{\eta}$ are the Nambu spinor wave-functions and energies for
quasiparticles, respectively. Within mean-field, the order parameter
takes the form $\Delta({\bf r})=-(U_{0}/2)\sum_{\eta}[u_{\uparrow\eta}v_{\downarrow\eta}^{*}f(E_{\eta})+u_{\downarrow\eta}v_{\uparrow\eta}^{*}f(-E_{\eta})]$
and, is to be solved self-consistently together with the atomic densities,
$n_{\sigma}\left({\bf r}\right)=(1/2)\sum_{\eta}[\left|u_{\sigma\eta}\right|^{2}f(E_{\eta})+\left|v_{\sigma\eta}\right|^{2}f(-E_{\eta})]$.
Here $f\left(x\right)\equiv1/\left(e^{x/k_{B}T}+1\right)$ is the
Fermi distribution function at temperature $T$. The chemical potential
$\mu$, implicit in ${\cal H}_{\sigma}^{S}({\bf r})$, can be determined
by the total number of atoms $N$ using the number equation $\int d{\bf r[}n_{\uparrow}\left({\bf r}\right)+n_{\downarrow}\left({\bf r}\right)]=N$.
As the impurity is placed at origin $r=0$, the BdG Hamiltonian preserves
rotational symmetry. Therefore, we take $\Delta({\bf r})=\Delta(r)$
and decouple the BdG equation into different angular momentum channels
indexed by an integer $m$, with which the quasiparticle wave functions
become, $[u_{\uparrow\eta}(r),u_{\downarrow\eta}(r)e^{i\varphi},v_{\uparrow\eta}(r)e^{i\varphi},v_{\downarrow\eta}(r)]e^{im\varphi}/\sqrt{2\pi}$.
By expanding $u_{\sigma\eta}\left(r\right)$ and $v_{\sigma\eta}\left(r\right)$
in the basis of 2D harmonic oscillators, the solution of BdG equation
converts to a matrix diagonalization problem. Numerically we have
to truncate the summation over energy levels $\eta$. This is done
by introducing a high energy cut-off $E_{c}$, above which a local
density approximation is used for high-lying wave-functions \cite{Liu2007}.
We have checked that such a hybrid procedure is numerically very efficient.

For the results presented here, we have solved self-consistently the
BdG equation for a cloud with $N=400$ atoms at zero temperature.
In 2D harmonic traps, it is convenient to use the Fermi radius $r_{F}=(4N)^{1/4}\sqrt{\hbar/(M\omega)}$
and Fermi energy $E_{F}=\hbar^{2}k_{F}^{2}/(2M)=\sqrt{N}\hbar\omega$
as the units for length and energy, respectively. The strength of
impurity scattering potential $V_{{\rm imp}}^{\sigma}$ will be measured
in units of $r_{F}^{2}E_{F}$. We have taken an interaction parameter
$E_{a}=0.2E_{F}$ and a spin-orbit coupling strength $\lambda k_{F}/E_{F}=1$.
With these parameters, the whole Fermi cloud becomes a topological
superfluid when the Zeeman field is larger than a threshold $h_{c}\simeq0.57E_{F}$
\cite{Liu2012a}. Let us first consider the localized impurities with
a delta-like scattering potential $V_{{\rm imp}}^{\sigma}\delta(r)$.

\textit{Emergence of universal impurity bound state}. --- According
to Anderson's theorem \cite{Anderson1959}, a conventional \textit{s}-wave
superfluid can barely be affected by non-magnetic impurities. In contrast,
magnetic impurities can break time-reversal symmetry of superfluid
and act as pair breakers, leading to the appearance of mid-gap state
- the so-called Yu-Shiba state - which is bound to localized impurities
inside the pairing gap \cite{Yu1965,Shiba1968}. The energy of such
a mid-gap bound state is determined by the strength of the impurity
scattering potential $V_{{\rm imp}}$. As $V_{{\rm imp}}$ increases,
the Yu-Shiba state moves from the upper gap edge to the lower gap
edge for the spin-up atoms and moves oppositely for the spin-down
atoms. In the presence of Rashba spin-orbit coupling, we have confirmed
numerically that the above statements continue to hold, even under
a Zeeman field, if the Fermi cloud is {\em not} a topological superfluid.
For a typical parameter $h=0.2E_{F}$, with increasing the strength
of magnetic impurity, we find that the position of the Yu-Shiba state
moves very quickly from one gap edge to the other.

\begin{figure}[t]
\begin{centering}
\includegraphics[clip,width=0.45\textwidth]{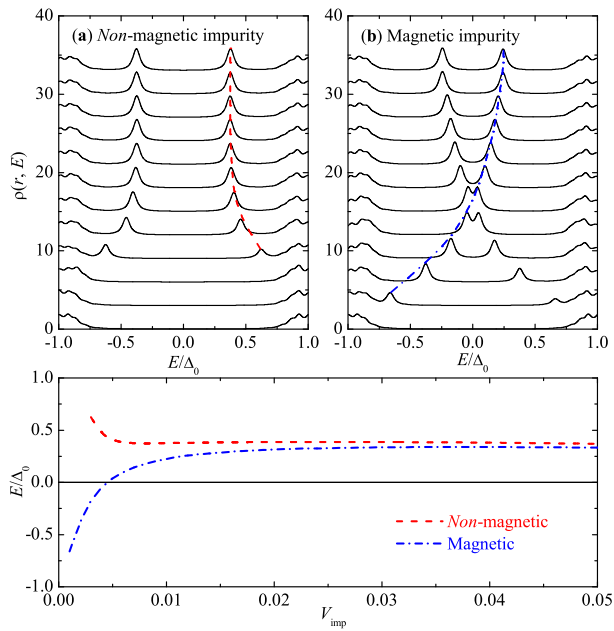} 
\par\end{centering}

\caption{(color online) Bound states induced by a non-magnetic delta-like impurity
(a) and by a magnetic delta-like impurity (b), $V_{{\rm imp}}^{\sigma}({\bf r})=V_{{\rm imp}}^{\sigma}\delta(r)$,
in a topological superfluid at $h=0.7E_{F}$, as shown by the peaks
in the total local density of state (LDOS) $\rho(r,E)$ at $k_{F}r=2$.
Here, $\rho(r,E)=\sum_{\sigma}\rho_{\sigma}({\bf r},E)$ and $\rho_{\sigma}({\bf r},E)=(1/2)\sum_{\eta}[\left|u_{\sigma\eta}\right|^{2}\delta(E-E_{\eta})+\left|v_{\sigma\eta}\right|^{2}\delta(E+E_{\eta})]$.
The dashed and dash-dotted lines highlight the resonance peak position
or the energy of bound states. From bottom to top, the impurity strength
increases from $V_{{\rm imp}}=0$ to $V_{{\rm imp}}=0.011r_{F}^{2}E_{F}$,
in steps of $0.001r_{F}^{2}E_{F}$. The curves are offset for clarity,
except for the lowest curve at $V_{{\rm imp}}=0$. (c) The energy
of bound states as a function of the impurity strength, in units of
the gap parameter at the trap center in the absence of impurity, $\Delta_{0}\simeq0.307E_{F}$.}

\label{fig1} 
\end{figure}

In contrast, once the Zeeman field is beyond the threshold $h_{c}$
so that the whole Fermi cloud becomes a topological superfluid, we
observe entirely different behavior, as revealed in Fig. 1. For non-magnetic
impurities, an unexpected bound state appears from one gap edge as
the impurity strength is larger than a critical strength $V_{{\rm imp}}\gtrsim0.004r_{F}^{2}E_{F}$.
As $V_{{\rm imp}}$ increases, the bound state moves towards, but
never reaches zero energy. In fact, its energy saturates quickly to
$E\simeq0.11E_{F}\simeq\Delta_{0}^{2}/E_{F}$, where $\Delta_{0}\simeq0.307E_{F}$
is the gap parameter at the trap center in the absence of impurity.
For magnetic impurities, the dependence of the position of the Yu-Shiba
state on the impurity strength is also strongly modified: at large
impurity scattering, the Yu-Shiba state now moves to $E\simeq\Delta_{0}^{2}/E_{F}$,
nearly at the {\em same} energy as the new bound state induced
by strong non-magnetic impurities. This coincidence in the energy
of bound states clearly indicates that in topological superfluids
a {\em universal} bound state emerges in the limit of strong impurity
scattering.

\begin{figure}[t]
\begin{centering}
\includegraphics[clip,width=0.45\textwidth]{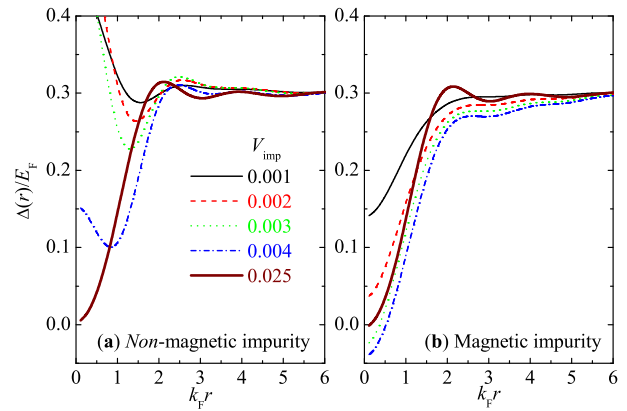} 
\par\end{centering}

\caption{(color online) Gap parameter as a function of impurity strength $V_{{\rm imp}}$
(in units of $r_{F}^{2}E_{F}$), for a non-magnetic impurity (a) and
for a magnetic impurity. In the limit of strong impurity scattering,
the gap parameter has the same spatial distribution, no matter the
impurity is non-magnetic or magnetic.}

\label{fig2} 
\end{figure}

\textit{Origin of the universal state}. --- The appearance of bound
states implies that the gap parameter would be strongly depleted close
to the impurity. In Fig. 2, we examine the spatial profile of order
parameter near the impurity. For a weak non-magnetic impurity, as
shown in Fig. 2(a), the gap parameter is already strongly modified
at $V_{{\rm imp}}\gtrsim0.004r_{F}^{2}E_{F}$. Being regarded as a
scattering potential for Bogoliubov quasiparticles \cite{CdGM}, the
gap parameter hence starts to accommodate a bound state. For a weak
magnetic impurity (Fig. 2(b)), the pair-breaking effect is always
significant enough to induce a Yu-Shiba bound state, as anticipated.
In the strong scattering limit, it is remarkable that the gap parameter
acquires a {\em universal} spatial profile, despite the type and
strength of impurities. It is fully depleted at the impurity site
and has a very similar distribution as the gap parameter inside a
vortex core. Therefore, we anticipate that the observed universal
bound state would resemble the well-known CdGM vortex-core bound states
\cite{CdGM}. Indeed, the energy of the universal impurity state,
$E\simeq\Delta_{0}^{2}/E_{F}$, is at the same order as that of CdGM
bound states.

Now, the formation of the universal bound state can be easily understood
from its analogy with the CdGM vortex-core state. As the gap parameter
is fully suppressed at the impurity site, we have a local point defect
(i.e., vacuum) that is topologically trivial. Due to the topological
nature of the Fermi cloud away from the impurity, there would be an
interface between the non-topological and topological components,
which can host a {\em gapless} Majorana edge state \cite{Shen2011}.
The observed universal impurity state is precisely such a Majorana
edge mode. However, its energy is not exactly zero due to the finite
confinement of the system \cite{Alicea2012}. As derived analytically
by Stone and Roy \cite{Stone2004} (see also Ref. \cite{Alicea2012}),
the dispersion relation of edge states in topological superfluids
with a confinement length $\xi$ is given by $E(m)=-(m+1/2)\Delta_{0}/(k_{F}\xi)$.
By assuming a characteristic length $\xi\sim\hbar v_{F}/\Delta_{0}$
for the gap parameter distribution \cite{CdGM}, where $v_{F}$ is
the Fermi velocity, we estimate that $E\sim\Delta_{0}^{2}/E_{F}$,
in good agreement with the observed energy of the universal bound
state.

\begin{figure}[t]
\begin{centering}
\includegraphics[clip,width=0.45\textwidth]{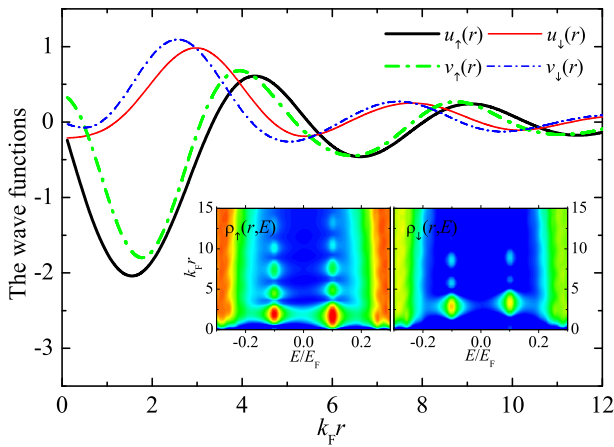} 
\par\end{centering}

\caption{(color online) The wave-function of the universal bound state, induced
by either non-magnetic or magnetic impurity in the strong scattering
limit. The inset shows the linear contour of LDOS for spin-up and
spin-down atoms near the impurity site.}

\label{fig3} 
\end{figure}

In Fig. 3, we examine the wave-function of the universal bound state.
Indeed, it satisfies approximately the symmetry $u_{\sigma}\left({\bf r}\right)=v_{\sigma}^{*}\left({\bf r}\right)$,
which should be obeyed by zero-energy Majorana fermions. In the inset,
we present the LDOS close to the impurity site. The universal bound
state is clearly visible within the gap. Experimentally, the LDOS
may be measured through spatially resolved radio--frequency (rf) spectroscopy
\cite{Shin2007}, which provides a cold-atom analog of the widely
used scanning tunneling microscope in solid state \cite{OurImpurityPRA}.
The wave-function of the universal bound state can therefore be determined
from the real-space structure of LDOS within the gap.

\begin{figure}[t]
\begin{centering}
\includegraphics[clip,width=0.45\textwidth]{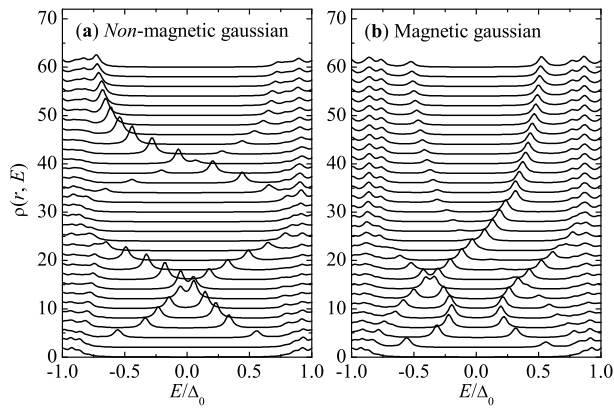} 
\par\end{centering}

\caption{Loss of the universal bound state for an extended impurity. Here,
we take a gaussian-shape scattering potential, $V_{{\rm imp}}^{\sigma}({\bf r})=[V_{{\rm imp}}^{\sigma}/(\pi d^{2})]\exp[-r^{2}/d^{2}]$,
with width $k_{F}d=0.5$. From bottom to top, the impurity strength
increases from $V_{{\rm imp}}=0$ to $V_{{\rm imp}}=0.06r_{F}^{2}E_{F}$,
in steps of $0.002r_{F}^{2}E_{F}$. Other parameters are the same
as in Fig. 1.}

\label{fig4} 
\end{figure}

\textit{Loss of university}. --- The universality of the impurity-induced
bound state can be lost if the impurity scattering has a finite width.
In this case, a hole will be created in the strong impurity scattering
limit, instead of a point defect. Therefore, there are a series of
edge states. The wave-function and energy of these edge states would
depend critically on the shape and strength of the impurity potential.
In Fig. 4, we show the bound states induced by a non-magnetic (a)
and a magnetic (b) gaussian impurity, with a finite width $k_{F}d=0.5$.
It is readily seen that with increasing the impurity strength the
bound state never approaches to a universal limit. We have checked
that for larger widths, the LDOS becomes very complicated, as more
and more bound states appear.

\textit{Experimental proposal}. --- We now show that ultracold Fermi
gases of $^{40}$K atoms is a potential candidate for observing the
predicted universal impurity-induced bound state. A 3D spin-orbit
coupled $^{40}$K Fermi gas was recently realized at Shanxi university
\cite{exptShanXi}. By loading a pancake-like optical trap $V({\bf r},z)=M[\omega^{2}r^{2}+\omega_{z}^{2}z^{2}]/2$
with trapping frequencies $\omega_{z}\gg\omega$ \cite{Dyke2011}
or using a deep 1D optical lattice \cite{Frohlich2011}, a 2D topological
superfluid with number of atoms $N\sim1000$ and size $r_{F}\sim100\mu m$
may be prepared at the temperature about $10nK$. It is convenient
to create the delta-like impurity potential by using a dimple laser
beam that has a sufficiently narrow beam width $d<1\mu m$ \cite{DimpleImpurity},
so that $k_{F}d\ll1$. By suitably tuning its frequency, the scattering
potential caused by the laser beam can be attractive or repulsive
for different spins. Thus, both non-magnetic and magnetic impurities
can be simulated. The resulting universal bound state may be visualized
by using the standard tool of spatially resolved rf-spectroscopy.
All the techniques required to observe the predicted universal state
are therefore within the reach of current experiments.

\textit{Application to other solid state systems}. --- Our results
are apparently applicable to the triplet superconductor Sr$_{2}$RuO$_{4}$.
For the possible 1D topological superconductor reported recently in
InSb nanowires \cite{Mourik2012}, a strong impurity potential would
split the 1D topological superconductor into two. Therefore, at the
impurity site we anticipate two universal bound states, with precise
zero-energy. The observation of such a pair of zero-energy Majorana
fermions is an unambiguous identification of the topological nature
of InSb nanowires.

\textit{Conclusion}. --- We have investigated the non-magnetic and
magnetic impurity scattering in an atomic topological superfluid and
have predicted the existence of universal bound state for strong impurity
scatterings. The observation of such a universal bound state - via
spatially resolved radio-frequency spectroscopy - is a smoking-gun
proof of atomic topological superfluidity. Our prediction seems within
experimental reach and opens the way to unambiguously characterizing
the topological properties of other solid-state systems, such as the
unconventional superconductor Sr$_{2}$RuO$_{4}$ and 1D topological
superconductor of InSb nanowires.

\textit{Note added}. --- After completing this work, we were aware
a related non-self-consistent \textit{T}-matrix calculation in 1D
topological superconductors, which predicted a bound state induced
by non-magnetic impurities \cite{Sau2012}. 
\begin{acknowledgments}
L.J. acknowledges stimulating discussions with Eite Tiesinga. H.H.
and X.-J.L are supported by the ARC DP0984522 and DP0984637 and the
NFRP-China 2011CB921502. H.P. acknowledges the support from the NSF,
the Welch Foundation (Grant No. C-1669) and the DARPA OLE program.
C.Y. is supported by the NSFC-China and the State Key Programs of
China.\end{acknowledgments}

\end{document}